\documentclass{PoS}
\usepackage{url}
\usepackage{siunitx}
\usepackage{nuc}
\usepackage{mathtools}
\usepackage{subfigure}

\usepackage{tikz}
\usetikzlibrary{shapes.multipart}
\usetikzlibrary{arrows}
\usetikzlibrary{shadows}
\usetikzlibrary{automata}
\usetikzlibrary{positioning}
\usepgflibrary{decorations.shapes}
\def\Put(#1,#2)#3{\leavevmode\makebox(0,0){\put(#1,#2){#3}}}
\usepackage{tikz-3dplot}
\usetikzlibrary{math} %needed tikz library
\usetikzlibrary{calc}
\usepgflibrary{arrows}
\tikzset{
    partial ellipse/.style args={#1:#2:#3}{
        insert path={+ (#1:#3) arc (#1:#2:#3)}
    }
}

\newcommand{\dd}{\text{d}}

\title{Electric dipole moment searches using storage rings}

\ShortTitle{Electric dipole moment searches using storage ring}
\author{\speaker{Frank Rathmann}\thanks{for the JEDI collaboration \protect\url{http://collaborations.fz-juelich.de/ikp/jedi/} (J\"ulich Electric Dipole moment Investigations)}\\
        Institute for Nuclear Physics (IKP), Forschungszentrum J{\"u}lich GmbH, 52428 J{\"u}lich, Germany
        E-mail: \email{f.rathmann@fz-juelich.de}}
\author{Nikolai N.\ Nikolaev\\
        L.D. Landau Institute for Theoretical Physics, 142432 Chernogolovka, Russia\\
        E-mail: \email{nikolaev@itp.ac.ru}}
\abstract{The Standard Model (SM) of Particle Physics is not capable to account for the apparent matter-antimatter asymmetry of our Universe. Physics beyond the SM is required and is either probed by employing highest energies (e.g., at LHC), or by striving for ultimate precision and sensitivity (e.g., in the search for electric dipole moments). Permanent electric dipole moments (EDMs) of particles violate both time reversal $(T)$ and parity $(P)$ invariance, and are via the $CPT$-theorem also $CP$-violating. Finding an EDM would be a strong indication for physics beyond the SM, and pushing upper limits further provides crucial tests for any corresponding theo\-retical model, e.g., SUSY. 

Up to now, EDM searches focused on neutral systems (neutrons, atoms, and molecules). Storage rings, however,  offer the possibility to measure EDMs of charged particles by observing the influence of the EDM on the spin motion in the ring. Direct searches of proton and deuteron EDMs, however, bear the potential to reach sensitivities beyond $\SI{e-29}{e.cm}$. Since the Cooler Synchrotron COSY at the Forschungszentrum J\"ulich provides polarized protons and deuterons up to momenta of 3.7 GeV/c, it constitutes an ideal testing ground and starting point for such an experimental program.

Besides the discussion of the achievements of the JEDI collaboration, and the description of an effort to perform a first direct deuteron EDM measurement at COSY, the report highlights in addition future technical developments that will pave the way toward EDM searches in dedicated rings. A recent advancement that grew out of the successful work performed by JEDI is the formation of the CPEDM Collaboration\protect\footnote{Charged Particle Electric Dipole Moment Collaboration \protect\url{http://pbc.web.cern.ch/edm/edm-default.htm}}, which aims at the design of an EDM prototype ring that could be hosted either at CERN or at COSY, will be discussed as well.}
\FullConference{23rd International Spin Physics Symposium - SPIN2018 -\\
		10-14 September, 2018\\
		Ferrara, Italy}
\begin{document}
%--------------------------------------------------------------------------------------------------------------------------------

%----------------------------------------------------------------------------------------------------------------------------
\section{Introduction}
%-------------------------------------------------------------------------------------------------------------------------
Electric dipole moments (EDMs) are one of the keys to understand the origin of and the baryogenesis in our Universe. In 1967 Andrei Sakharov formulated three conditions for baryogenesis~\cite{Sakharov:1967dj}: 
\begin{enumerate}
 \item Early in the evolution of the Universe, the baryon number conservation must be violated sufficiently strongly.
 \item The $C$ and $CP$ invariances, and $T$ invariance thereof,  must be violated.
 \item At the point in time when the baryon number is generated, the evolution of the Universe must be out of thermal equilibrium.
\end{enumerate}
$CP$ violation in kaon decays is known since 1964, it has been observed in  $B$-decays and in  charmed meson decays, and based on the existing data can be described by the $CP$-violating phase in the Cabibbo-Kobayashi-Maskawa matrix~\cite{PhysRevLett.110.101802,Polycarpo:2012yh}). $CP$ and $P$ violation entail  non-vanishing $P$ and $T$ violating EDMs of elementary particles. Although extremely successful in many aspects, the Standard Cosmological Model (SCM) has one pronounced weaknesses; it fails miserably in the expected baryogenesis rate. %the CKM matrix of the SM amounts to the relevant effective coupling $\leq 10^{-20}$~\cite{doi:10.1146/annurev.nucl.49.1.35}, much too feeble to account for the observed baryon density. 
The observed baryon asymmetry $\eta$ of the Universe is expressed via 
\begin{equation}
 \eta = \displaystyle (n_b - n_{\bar{b}})/n_\gamma\,,
\end{equation}
where $n_b$ and $n_{\bar{b}}$ denote the number of baryons and anti-baryons, and $n_\gamma$ the number of relic photons. The discrepancy between observation and expectation from the Standard Cosmological Model (SCM) amounts to about 9 orders of magnitude (see Table\ref{table:1}).
\begin{table}[hbt]
\renewcommand*{\arraystretch}{1.3}\small
\begin{tabular}{l |l|l}
                       & $\eta = \displaystyle (n_b - n_{\bar{b}})/n_\gamma$ & \\\hline
 Observation              & $\left( 6.11^{+ 0.3}_{-0.2} \right) \times \num{e-10} $      &  Best Fit Cosmological Model\,\cite{Bennett:2003bz}\\
                       & $(5.53-6.76) \times \num{e-10}$ & WMAP\,\cite{Barger:2003zg} \\
 Expectation from SCM  & $\sim \num{e-18}$ & Bernreuther (2002)\,\cite{Bernreuther:2002uj}\\
\end{tabular}
\caption{\label{table:1}Observation and expectation from Standard Cosmological Model (SCM).}
\end{table}

Simultaneously, the SM predicts exceedingly small electric dipole moments of nucleons  $ \num{e-33} < d_n < \SI{e-31}{e.cm}$\,\cite{Khriplovich:1997ga}, way below the current upper bound for the neutron EDM, which is $d_n \lessapprox \SI{2.9e-26}{e.cm}$\,\cite{Afach:2015sja}, and also beyond the reach of future EDM searches\,\cite{roberts-marciano:2010}. 

In the quest for physics beyond the SM one could follow either the high energy trail or look into new methods which offer very high precision and sensitivity. Supersymmetry is one of the most attractive extensions of the SM. The SUSY predictions span typically a range of $10^{-29} < d_n < 10^{-24}$ e$\cdot$cm  and precisely this range is targeted in the new generation of EDM searches~\cite{roberts-marciano:2010}, discussed here.

There is  consensus among theorists that measuring  the EDM of the proton, deuteron and helion is as important as that of the neutron\,\cite{Bsaisou:2015:1,Bsaisou:2015:2}. The EDMs could have a non-trivial isospin dependence and $d_d \neq d_p + d_n$, even if the $CP$-violation comes from the isoscalar QCD $\theta$-term\,\cite{Guo:2012vf}. Furthermore, it has been argued some 25 years ago that $T$-violating nuclear forces could substantially enhance nuclear EDMs\,\cite{Flambaum:1985gv,Flambaum:1985ty}. At the moment, there are no significant {\it directly} determined upper bounds available on $d_e$, $d_p$ and $d_d$. The current status of EDM searches is reflected in Table\,\ref{table:2}.
\begin{table}[hbt]
 \begin{tabular}{r|r|l|l|l}	
  \textbf{Particle}   & \textbf{Current  limit}              & \textbf{Goal}                                  & $\mathbf{d_n}$ \textbf{equivalent goal }                        & \textbf{Date [ref]}  \\\hline
  Electron   & $< \num{8.7e-29}$          & $\approx \num{e-29}$                  &                                               & 2014\,\cite{ACME2014}\\
  Muon       & $< \num{1.8e-19}$          &                                       &                                               &       2009 \cite{PhysRevD.80.052008}\\
  Tau        & $< \num{1e-17}$            &                                       &                                               &
  2003 \cite{INAMI200316}\\\hline
  Lambda     & $< \num{3e-17}$                            &                                       &                               & 
  1981 \cite{PhysRevD.23.814}\\
  Neutron    & $(-0.21 \pm 1.82)\times\num{e-26}$  & $\approx \num{e-28}$                  & $\num{e-28}$                                  & 2015 \cite{Afach:2015sja}\\
  $\Hg{199}$ & $< \num{7.4e-30}$           & $\num{e-30}$                          & $<\num{1.6e-26}$\,\cite{PhysRevLett.91.212303} & 2016 \cite{PhysRevLett.116.161601}\\
  $\Xe{129}$ & $< \num{6.0e-27}$           & $\approx \num{e-30}$ to $\num{e-33}$  & $\approx \num{e-26}$ to  $\num{e-29}$     & 2001 \cite{PhysRevLett.86.22}\\   
  Proton     & $< \num{2e-25}$           & $\approx \num{e-29}$                  & $\num{e-29}$                              & 2016 \cite{PhysRevLett.116.161601}\\
  Deuteron   &    not available yet        & $\approx \num{e-29}$                  & $\approx \num{3 e -29}$ to $ \num{5e-31}$ & \\\hline
    \end{tabular}
    \caption{\label{table:2} Current limits, goals and $d_n$ equivalent goals for various particles.}
\end{table}

\section{Charged particle EDM searches using storage rings}
\subsection{Experimental requirements}

The experimental requirements for charged particle EDM searches using storage rings are very demanding and require the development of a new class of high-precision, primarily electric storage rings. Precise alignment, stability, field homogeneity, and shielding from perturbing magnetic fields play a crucial role. Beam intensities around $N = \num{4e10}$ particles per fill with a polarization of $P=0.8$ are anticipated. Large electric fields of $E = \SI{10}{MV/m}$ and   long spin coherence times of about $\tau_\text{SCT} = \SI{1000}{s}$ are necessary.  Efficient polarimetry with large analyzing power of $A_y \simeq 0.6$, and high efficiency of detection $f \simeq 0.005$ need to be provided.
In terms of the above numbers, this would lead to statistical uncertainties of 
    \begin{equation}
    \sigma_\text{stat} = \frac{2\hbar}{\sqrt{N \, f} \, \tau_\text{SCT} \, P \, A_y \, E }  \quad \Rightarrow \quad \sigma_\text{stat}(\SI{1}{yr}) = \SI{1.9e-29}{e.cm}\,,
   \label{eq:statistical-error-EDM-experiment}
    \end{equation}
where for one year of data taking \num{10000} cycles of \SI{1000}{s} duration is assumed. The experimentalist's goal must be to provide systematic uncertainties $\sigma_\text{syst}$ to the same level.

\subsection{Spin precession in a storage ring and frozen-spin method}

In the rest frame of the particle in a storage ring, the equation of motion for the spin vector $\vec S$ in the presence of an electric field $\vec E$ and magnetic field $\vec B$ can be written as
\begin{equation}
 \frac{\dd \vec S}{\dd t} = \vec \Omega \times \vec S = \vec \mu \times \vec B + \vec d \times \vec E\,,
 \label{eq:rest-frame-spin-precession}
\end{equation}
where $\mu$ denotes the magnetic moment, and $d$ the electric dipole moment. The spin precession frequency of a particle on the closed orbit due to its magnetic dipole moment (MDM) \textit{relative} to the direction of flight  can be expressed as 
\begin{equation}
\begin{split}
 \vec \Omega & = \vec \Omega_\text{MDM} - \vec \Omega_\text{cyc} \\
             & = - \frac{q}{\gamma m} \left[G \gamma \vec B_\perp + (1 + G) \vec B_\parallel - \left(G\gamma - \frac{\gamma}{\gamma^2 -1} \right) \frac{\vec \beta \times \vec E}{c}\right]\,.  
\end{split}
\end{equation}
$\vec \Omega = 0$ is called \textit{frozen spin}, because in this case momentum and spin stay aligned. In the absence of magnetic fields ($B_\perp = \vec B_\parallel = 0$), 
\begin{equation}
 \vec \Omega=0, \text{ if } \left(G\gamma - \frac{\gamma}{\gamma^2 -1} \right) = 0.
 \label{eq:magic-momentum}
\end{equation}
This can be realized only for particles with $G>0$, such as proton ($G_p=1.793$) or electron ($G_e=0.001$). 
For protons, Eq.\,(\ref{eq:magic-momentum}) leads to the so-called \textit{magic momentum} $p_\text{magic}$
  \begin{equation}
      G_p - \frac{1}{\gamma^2-1} = 0 \quad \Leftrightarrow \quad G_p = \frac{m^2}{p_\text{magic}^2} 
      \quad \Rightarrow \quad p_\text{magic} = \frac{m}{\sqrt{G_p}} = \SI{700.740}{MeV.c^{-1}}\,.
  \end{equation}

Storing protons in a ring with purely electrical deflection elements at magic momentum freezes the horizontal spin precession, \textit{i.e.}, 
the proton spins remain  aligned along the direction of flight.   In a purely electric machine with $\vec B=0$, Eq.\,(\ref{eq:rest-frame-spin-precession}) then implies the development of a vertical polarization component $p_y(t)$. The derivative of which is proportional to the electric dipole moment. Here it should be noted that freezing the spin precession works for any spin orientation. Obviously, the highest sensitivities can be reached when $\vec d$ and $\vec E$ are orthogonal, hence when $\vec d$ points along the momentum.

Magic machines for light ions with frozen spin can be envisioned to allow for a measurement using different particle types. The general solution for the ratio of outward electric field $E_x$ to the vertical magnetic field $B_y$ fulfilling the magic condition, derived from the Thomas-BMT equation, can be expressed as
    \begin{equation}
         \frac{E_x}{B_y} = \frac{G  c \beta \gamma^2}{1 - G \beta^2 \gamma^2}\,, 
               \label{eq:frozen-spin-with-magnetic-field}
    \end{equation}
(right-handed coordinate system, with $z$ along beam direction). Equating the Lorentz force and the relativistic centrifugal force, yields then for a specific radius the required electric and magnetic fields. The required parameters for electric and magnetic field for a circular machine with radius $r = \SI{25}{m}$ are listed in Table\,\ref{table:3}.
\begin{table}[hbt]
   \renewcommand*{\arraystretch}{1.2}
   \begin{center}
    \begin{tabular}{lrrrrr}
particle  & $G$      & $p$\,$[\si{MeV.c^{-1}}]$   & $T$\,$[\si{MeV}]$ & $E_x$\,$[\si{MV.m^{-1}}]$ & $B_y$\,$[\si{T}]$   \\ \hline
proton    & $1.793$  & $\num{700.740}$            & $\num{232.792}$   & $\num{16.772}$          & $0.000$           \\
deuteron  & $-0.143$ & $\num{1000.000}$           & $\num{249.928}$   & $\num{-4.032}$          & $\num{0.162}$          \\
helion    & $-4.184$ & $\num{1200.000}$           & $\num{245.633}$   & $\num{14.654}$          & $\num{-0.044}$       \\ \hline 
    \end{tabular}
   \end{center}
   \caption{\label{table:3} Example for frozen spin conditions for protons, deuterons and helions with and without magnetic fields for a circular machine with radius $r = \SI{25.000}{m}$ using Eq.\,(\ref{eq:frozen-spin-with-magnetic-field}).}
\end{table}

Measurement of the EDM of protons, deuterons and helions can be anticipated to take place in one and the same machine.

%--------------------------------------------------------------------------------------------------------------------------
\section{Progress toward storage ring EDM experiments}
\begin{figure}
\begin{center}
   \includegraphics[width=0.7\textwidth]{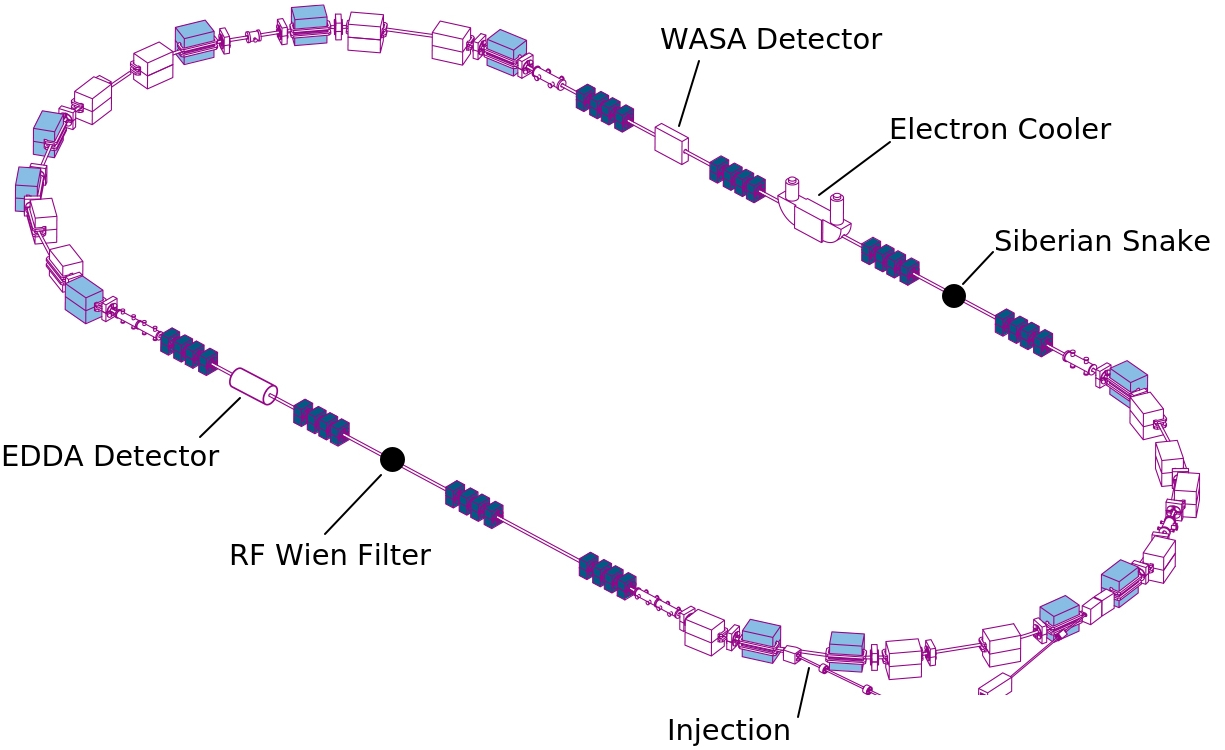}
\end{center}
\caption{\label{fig:COSY-landscape}Landscape of COSY with the main installations employed to perform a first direct measurement of the deuteron EDM.}
\end{figure}

%---------------------------------------------------------------------------------------------------------------------------
The COoler SYnchrotron COSY has been formerly used as spin-physics machine for hadron physics experiment. It provides phase-space cooled internal and extracted beams of polarized protons and deuterons at momenta of $p = 0.3$ to $\SI{3.7}{GeV/c}$. Since about 2012,  COSY is heavily used to complement the spin-physics tool box for storage ring EDM experiments, as it provides an ideal starting point for accelerator related R\&D. In addition, as will be outlined below, COSY will be used to carry out a first direct measurement of deuteron EDM. Figure\,\ref{fig:COSY-landscape} shows the main installations presently in use for this purpose at COSY.

\subsection{Precision determination of the spin tune }

The JEDI collaboration developed a new technique to determine the spintune $\nu_s$ in a machine\,\cite{PhysRevLett.115.094801}.
The spin tune $\nu_s$ is determined to about $  \num{e-8}$ in a $\SI{2}{s}$ time interval, and in a $\SI{100}{s}$ cycle at $t\approx \SI{38}{s}$, the relative uncertainty of the spin tune amounts to $\Delta \nu_s/\nu_s \approx \num{e-10}$. With this, a new precision tool for accelerator physics has become available to study systematic effects in a storage ring, \textit{e.g.}, the long term stability of an accelerator.

\subsection{Optimization of spin-coherence time }
\label{sec:optimization-of-spin-coherence-time}
One of the main obstacles for any storage ring EDM experiment is the decoherence of the in-plane polarization. Using sextupole magnets to correct higher order effects, in 2014 at COSY spin coherence times (SCT) of about $\tau_\text{SCT} \approx \SI{400}{s}$ could be reached\,\cite{PhysRevSTAB.17.052803}. Since 2016, typical values routinely exceeding $\tau_\text{SCT} =   \SI{800}{s}$ are available\,\cite{PhysRevLett.117.054801}. This pronounced progress has not been anticipated. It should be emphasized that large spin coherence times are of particular importance, because $ \sigma_\text{stat} \propto {\tau_\text{SCT}}^{-1}$ (see Eq.\,(\ref{eq:statistical-error-EDM-experiment})).

\subsection{Phase locking the spin precession}
\label{sec:phase-locking-the-spin-precession}

In a machine with purely magnetic deflection and focusing like COSY, it is not possible to freeze the spins. Using an RF device that operates on a harmonic of the spin-precession frequency is the only possible approach toward an EDM measurement in COSY. In order to achieve a good precision for such a measurement, phase-locking is necessary, making sure that phase between the spin-precession and the device RF is maintained throughout the measurement. To this end, a feedback system has been developed that stabilizes the phase of the spin precession relative to the phase of an RF devices, providing a so-called \textit{phase-lock}. The feedback system maintains the resonance frequency, and the phase between spin precession and device RF (\textit{e.g.}, solenoid or Wien filter). As a major achievement, an error of the phase-lock of $\sigma_\phi = \SI{0.21}{rad}$ has been achieved\,\cite{PhysRevLett.119.014801,PhysRevAccelBeams.21.042002}.

In the presence of a long spin-coherence time, phase-locking of the in-plane polarization can be viewed as providing a co-magnetometer for the resonant buildup of a vertical polarization component using an RF Wien filter (cf. Sec.\,\ref{sec:precursor-experiment}).

\subsection{Spin tune mapping}
\label{sec:spin-tune-mapping}

Precision experiments, such as the search for electric dipole moments of charged particles using storage rings, demand for an understanding of the spin dynamics with unprecedented accuracy. As the ultimate aim is to measure the electric dipole moments with a sensitivity up to 15 orders in magnitude better than the magnetic dipole moment of the stored particles. For this reason, the background to the signal of the electric dipole  from rotations of the spins in spurious magnetic fields of the storage ring must be understood. One of the observables, especially sensitive to the imperfection magnetic fields in the ring is the angular orientation of stable spin axis. For the first time, the JEDI collaboration succeeded  to determine experimentally the stable spin axis. A new method called \textit{spin tune mapping} was developed, and the angular orientation of the stable spin axis at two different locations in the COSY ring has been determined to an unprecedented accuracy of better than $\SI{2.8}{\micro.rad}$\,\cite{PhysRevAccelBeams.20.072801}.

%----------------------------------------------------------------------------------------------------
\section{Technical challenges and developments}
%----------------------------------------------------------------------------------------------------
\subsection{Overview}

Charged particle EDM searches require development of a new class of high-precision machines with mainly electric fields for bending and focusing. Some of the technical challenges involved in this will be discussed in the following sections:
\begin{itemize}
  \item Spin coherence time $\tau_\text{SCT} \sim \SI{1000}{s}$ (see Sec.\,\ref{sec:optimization-of-spin-coherence-time}\,\cite{PhysRevLett.117.054801}).
 \item Large electric field gradients $\sim 10 \text{ to } \SI{20}{MV \per m}$ (see Sec.\,\ref{sec:deflector}).
 \item Beam position monitoring with precision of $\SI{10}{nm}$ (see Sec.\,\ref{sec:bpms}).
 \item Continuous polarimetry with relative errors $< \SI{1}{ppm}$\,\cite{Brantjes201249}. Analyzing power measurement to provide better data and a novel polarimeter design are discussed in Secs.\,\ref{sec:polarimetry-data-base} and \ref{sec:beam-polarimeter}.
 \item Magnetic imperfections (see Sec.\,\ref{sec:study-of-machine-imperfections}).
\item Prototype EDM storage ring (see Sec.\,\ref{sec:prototype-EDM-storage-ring}).
  \item Alignment of ring elements, ground motion, ring imperfections.
\item For deuteron EDM with frozen spin: precise reversal of magnetic fields for CW and CCW beams required.
\end{itemize}
 
\subsection{$E/B$ deflector development}
\label{sec:deflector}
 
In the framework of the CPEDM collaboration\footnote{Charged Particle Electric Dipole Moment Collaboration \url{http://pbc.web.cern.ch/edm/edm-default.htm}}, a prototype EDM storage ring is presently being developed (see Sec.\,\ref{sec:prototype-EDM-storage-ring}). In conjunction with this development, electrostatic deflector elements are being designed that provide radial electric fields. Combined elements that generate in addition vertical magnetic fields are being developed as well. 

The development takes place in two stages that are jointly organized by IKP of Forschungszentrum J\"ulich and RWTH Aachen University. In stage 1, a laboratory setup, developed at RWTH Aachen, employs scaled-down electrodes. The purpose of this investigation is to identify potential materials, coatings and surface treatment that can be applied in order to achieve high electric fields. With a \SI{30}{kV} power supply, and appropriately reduced distances of up to a few mm between the electrodes, large electric fields of interest can be achieved. First results using polished stainless steel electrodes are reported in\,\cite{Grigoryev:2018ipo}.

\begin{figure}[t]
\begin{center}
\subfigure[\label{fig:large-scale-deflector-setup:a}  $\SI{64}{t}$ dipole magnet. ]
{\includegraphics[height=0.26\textheight]{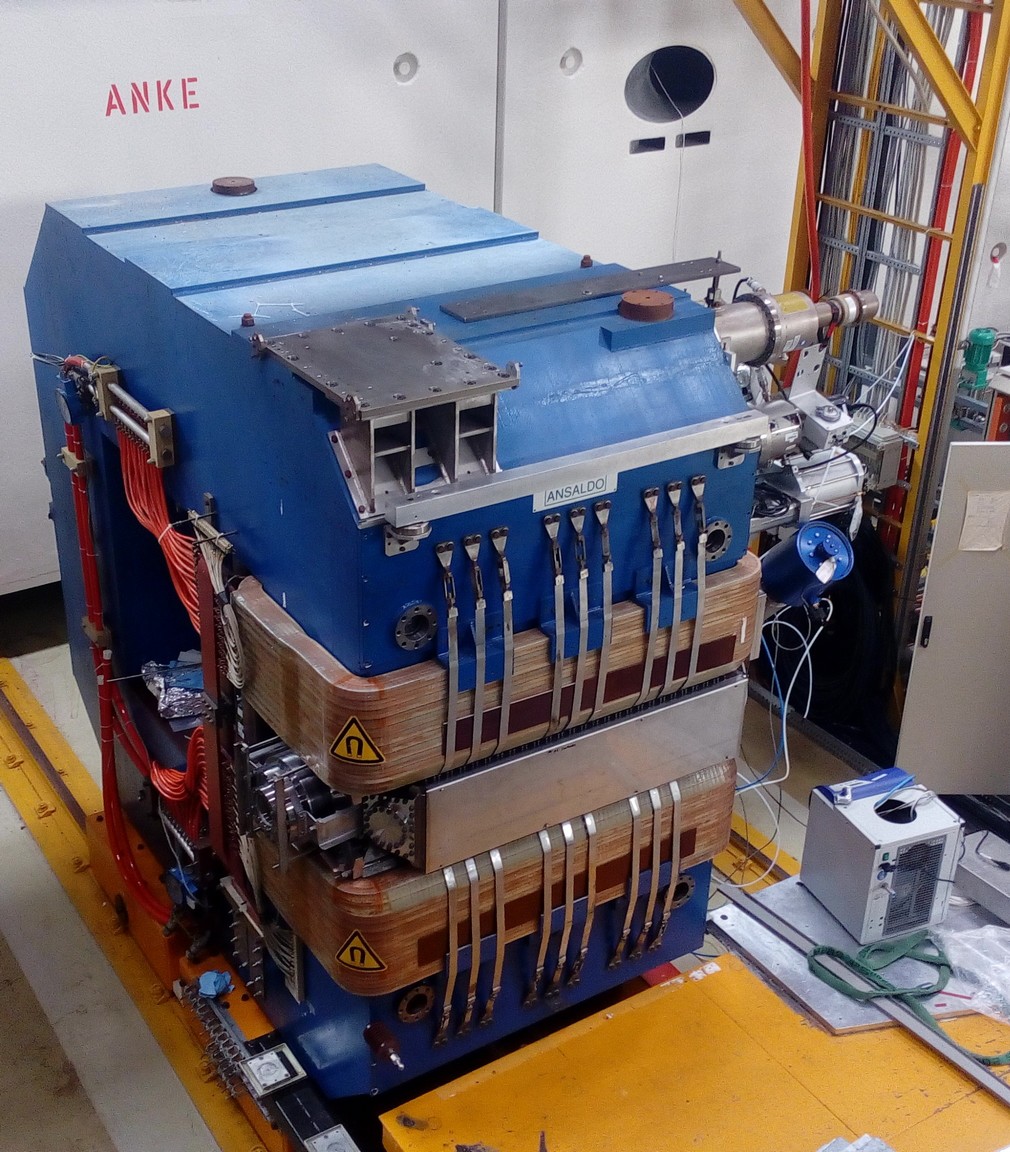}}
\hspace{0.2cm}
\subfigure[\label{fig:large-scale-deflector-setup:b} Setup to be installed in the magnet chamber.]
{\includegraphics[height=0.26\textheight]{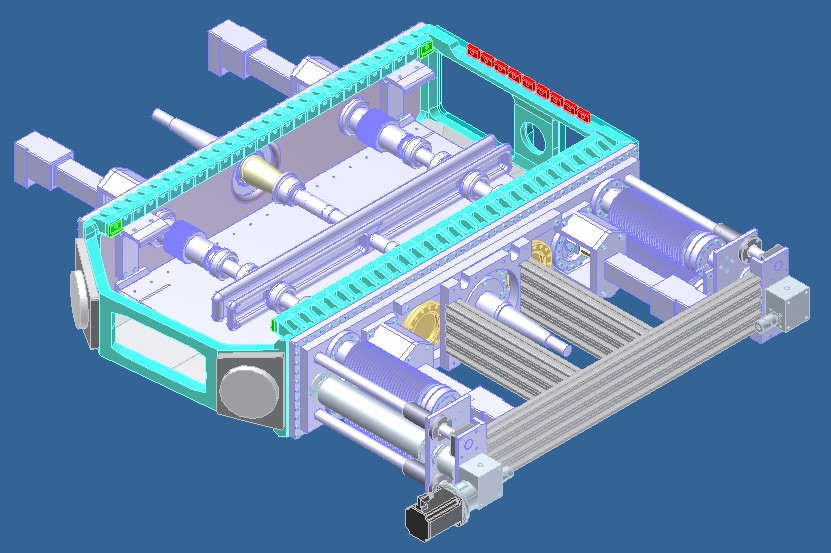}}\\
\end{center}
\caption{\label{fig:large-scale-deflector-setup}  The magnet can produce up to $B_\text{max} = \SI{1.6}{T}$ in a gap of height  $h_\text{g} = \SI{200} {mm}$. The electrode length is $\ell = \SI{1020}{mm}$, the electrode height $h_\text{e} = \SI{90}{mm}$, and the electrode spacing $S = 20 \text{ to } \SI{120}{mm}$. The maximum applied voltage field $U = \pm \SI{200}{MV}$. Foreseen material is aluminum coated by TiN.}
\end{figure}

Stage 2 of the deflector development program aims at tests with real-size deflector elements of a length of about $\ell = \SI{1000}{mm}$, employing two \SI{200}{kV} power supplies and plate distances ranging from 20 to \SI{120}{mm}. The experimental setup makes use of a large-gap spectrometer magnet, as shown in Fig.\,\ref{fig:large-scale-deflector-setup}. 

\subsection{Beam-position monitors}
\label{sec:bpms}
Storage ring EDM experiments require very precise orbit measurements along the circumference of the ring. The JEDI collaboration has begun to develop a new type of compact beam-position monitor based on a segmented Rogowski coil\,\cite{Trinkel:2017}.

\begin{figure}[bt]
\begin{center}
\subfigure[\label{fig:bpm:a} \small Conventional split-cylinder beam-position monitor with an installation length of $\approx \SI{20}{cm}$.]
{\includegraphics[height=0.25\textheight]{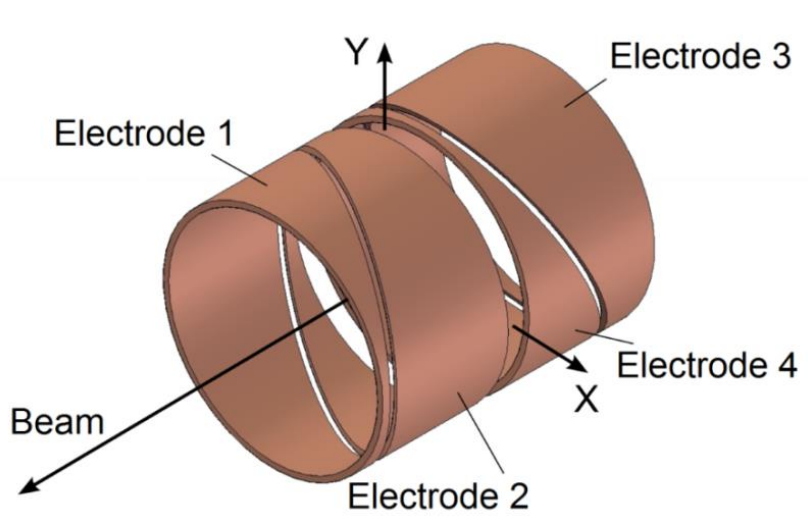}}
\hspace{0.2cm}
\subfigure[\label{fig:bpm:b} \small Rogowski pickup based on a segmented toroidal coil.]
{\includegraphics[height=0.25\textheight]{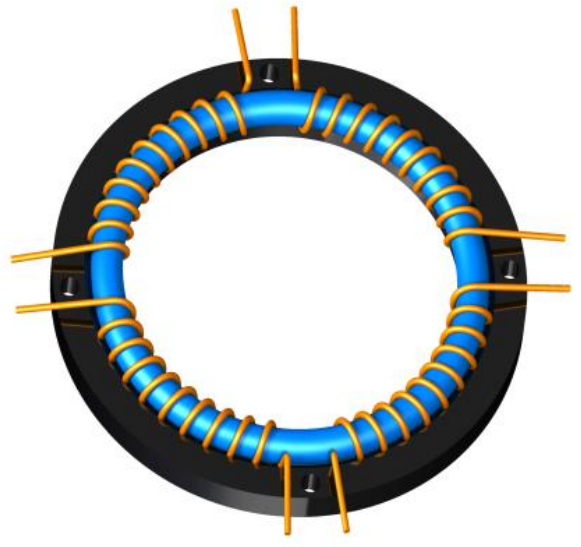}}\\
\end{center}
\caption{\label{fig:bpm} \small The main advantage of the Rogowski design is that with a toroid diameter of $d_\text{t} = \SI{10}{cm}$, and a coil diameter $d_\text{c} = \SI{1}{cm}$, the installation length is much smaller than the one of the split-cylinder design.}
\end{figure}

The main advantage of this design is a short installation length of $\approx \SI{1}{cm}$ (along the beam direction), while the sensitivity for bunched beam positions is estimated to be better than the conventional split-cylinder design, shown in Fig.\,\ref{fig:bpm}.

Two of these Rogowski pickups are already in operation in COSY at the entrance and exit of the RF Wien filter (WF)\,\cite{spin2018:Alexander.Nass}.

\subsection{$dC$ polarimetry data base}
\label{sec:polarimetry-data-base}
Due to the large analyzing power and differential cross section in the forward region, $dC$ elastic scattering constitutes a well-suited polarimeter reaction for deuteron EDM measurements. In order to provide precise input to Monte-Carlo simulations for an optimized beam polarimeter design, the analyzing powers and the differential cross sections were measured at six different deuteron beam kinetic energies in the range of \SI{170}{MeV} to \SI{380}{MeV}\,\cite{spin2018:Fabian.Mueller,Pretz2019}

\subsection{Beam polarimeter}
\label{sec:beam-polarimeter}
Up to now, the EDM-related COSY experiments, carried out by the JEDI collaboration, employed \textit{existing} detector installations as polarimeters (\textit{e.g.}, EDDA\, \cite{Albers:2004iw} and WASA\,\cite{Adam:2004ch,Bargholtz:2008aa}). A few years ago, the decision was taken to develop a high-precision beam polarimeter with an internal Carbon target based on LYSO scintillation material. 

This detector material, produced by Saint-Gobain Ceramics \& Plastics\footnote{\url{Saint-Gobain Crystals, https://www.crystals.saint-gobain.com}}, is a Cerium doped Lutetium based scintillation crystal, Lu$_{1.8}$Y$_{.2}$SiO$_5$:Ce. Compared to NaI, LYSO provides higher density (7.1 vs \SI{3.67}{g \per cm^3}), and a  very fast decay time (45 vs \SI{250}{ns})\,\cite{spin2018:Dito.Shergelashvili}. After several commissioning runs with external beam, the detector system will be installed at COSY in 2019. 

\subsection{Study of machine imperfections}
\label{sec:study-of-machine-imperfections}

JEDI developed a new method to investigate magnetic machine imperfections based on the highly accurate determination of the spin-tune. This \textit{spin-tune mapping} technique used the two available cooler solenoids of COSY as (makeshift) spin rotators to generate artificial imperfection fields. The measurement of the shifts of the spin tune as function of the spin kicks of the two solenoids yields the map\,\cite{PhysRevAccelBeams.20.072801, spin2018:Kolya.Nikolaev}. 

The location of the saddle point of the map determines the tilt of the stable spin axis caused by the magnetic imperfections. It is possible to control the background to the direction of the stable spin axis $\vec c$ from magnetic dipole moment rotations at a level $\Delta c = \SI{2.8e-6}{rad}$\,\cite{PhysRevAccelBeams.20.072801}. The systematics-limited sensitivity for a deuteron EDM measurement at COSY amounts to $\sigma_d \approx \SI{e-20}{e.cm}$.

\subsection{From JEDI to CPEDM: a prototype EDM storage ring}
\label{sec:prototype-EDM-storage-ring}
 
In view of the various technical challenges involved in building the final all-electric ring, as \textit{e.g.}, described in\,\cite{doi:10.1063/1.4967465}, as next step, the CPEDM collaboration decided to design and build a demonstrator ring for charged-particle EDM searches. The new CPEDM collaboration, which evolved out of the success and the achievements of JEDI, brings together scientists from CERN and the JEDI collaboration.  The project is part of the Physics Beyond Collider (PBC) process presently carried out at CERN, and the
European Strategy for Particle Physics Update. A possible host site for the prototype EDM storage ring is either COSY or CERN.

The scope of the project is to provide for protons at a kinetic energy of $T = \SI{30}{MeV}$ an all-electric machine operation with simultaneous clockwise (CW) and counter-clockwise (CCW) orbiting beams of the machine. The circumference of the machine is about \SI{100}{m}. At $T=\SI{45}{MeV}$ using vertical magnetic fields superimposed on the radial electric fields in the deflector elements, frozen-spin operation for protons shall be possible. Items to be studied with the prototype ring include:
\begin{itemize}
   \item Storage time investigations,
   \item CW/CCW operation.
   \item Spin coherence time studies.
   \item Polarimeter studies.
   \item Studies of magnetic moment effects due to imperfect shielding and artificially introduced magnetic fields.
   \item A direct measurement of the EDM of the proton.
   \item Tests of stochastic cooling.
 \end{itemize}
Further details about this project can be found in a contribution to these proceedings\,\cite{spin2018:Sig.Martin}.

\section{Proof of principle EDM (\textit{precursor}) experiment using COSY }
\label{sec:precursor-experiment}
Highest EDM sensitivity shall be achieved with a new type of machine, namely with an \textit{electrostatic} circular storage ring, where
the centripetal force is produced by electric fields. This $E$ field couples to the EDM of the orbiting particles and provides the desired  sensitivity ($< \SI{e-28}{e.cm}$). It is obvious that in such an environment, magnetic fields mean evil, since the MDM ($\mu_N = e\hbar/2m_N c \approx \SI{e-14}{e.cm}$) is vastly larger than the EDM we are after.
 
The idea behind such a proof-of-principle experiment (so-called precursor experiment) is to use a novel $\vec E \times \vec B$ RF Wien filter\,\cite{Rathmann:2013rqa,PhysRevSTAB.16.114001} to accumulate the EDM related spin rotation in order to make them measurable. In a magnetic machine, the particle spins precess about the local stable spin axis. In an ideal machine, this axis corresponds to the vertical ($y$) direction of the magnetic fields $\vec B_\text{dipole} \propto \vec e_y $ in dipole magnets. In this situation, an RF device that is operating on some harmonic of the spin-precession frequency can be used to accumulate the EDM effect as function of time in the cycle, provided the particle ensemble is coherently precessing in the horizontal plane (see Sec.\,\ref{sec:optimization-of-spin-coherence-time}).

In order for the RF system of the Wien filter to stay tuned precisely on a harmonic of the spin-precession frequency, a phase-lock between the spin-precession of the particle ensemble in the ring and the RF of the Wien filter is needed, as described in Sec.\,\ref{sec:phase-locking-the-spin-precession}. The horizontally precessing polarization serves as a co-magnetometer for the buildup of the vertical polarization (EDM) signal.  The goal of the experiment is to show that a conventional \textit{magnetic} storage ring can be employed to obtain a first direct EDM measurement of the deuteron (or proton).

\subsection{Technical realization and modeling of the RF Wien filter}
The technical realization and a report about the commissioning of the RF Wien filter\,\cite{Slim:2016pim} at COSY is available in these proceedings in Ref.\,\cite{spin2018:Alexander.Nass}. Two additional aspects of this development shall be mentioned here. 

Mechanical tolerances and misalignments decrease the simulated field quality of the RF Wien filter, and it is therefore important to consider them in the simulations. In particular, for the EDM measurement, it is important to quantify these field errors systematically. Since Monte-Carlo simulations are computationally very expensive, an efficient surrogate modeling scheme based on the Polynomial Chaos Expansion method to compute the field quality in the presence of tolerances and misalignments has been developed, which was subsequently used to perform a sensitivity analysis of the RF Wien filter at zero additional computational cost\,\cite{Slim:2017bic}.

We have developed an implementation of the polynomial chaos expansion as a fast solver of the equations of beam and spin motion of charged particles in electromagnetic fields, and it could be shown that, based on the stochastic Galerkin method\footnote{The Galerkin method\,\cite{Augustin2012} constitutes one of the many possible finite element method formulations that can  be used for discretization.}, this computational framework substantially reduces the required number of tracking calculations compared to the widely used Monte Carlo method\,\cite{Slim:2016dct}. 

\subsection{Model calculation of polarization evolution}
EDM induced vertical polarization oscillations in an experimental situation with an RF Wien filter  can generally be described by
\begin{equation}
 p_y(t)  = a \, \sin ( \Omega^{p_y} \, t + \phi_\text{RF}) \,.
 \label{eq:initial-slope-from-full-oscillation}
\end{equation}
The associated EDM resonance strength $\varepsilon^\text{EDM}$ can be defined as the ratio of angular frequency $\Omega^{p_y}$ relative to the orbital angular frequency $\Omega^\text{rev}$ in the machine,
\begin{equation}
 \varepsilon^\text{EDM} = \frac{\Omega^{p_y}}{\Omega^\text{rev}}\,.
 \label{eq:resonance-strength}
\end{equation}

The term ``EDM'' in Eq.\,(\ref{eq:resonance-strength}) applies to the case that only the EDM contributes to $\Omega^{p_y}$. In practice, the resonance strength will receive contributions from other sources, such as rotations of the RF Wien filter and solenoidal fields in the ring that generate unwanted spin kicks.

\begin{figure}[t]
\begin{center}
    \includegraphics[width=0.75\textwidth]{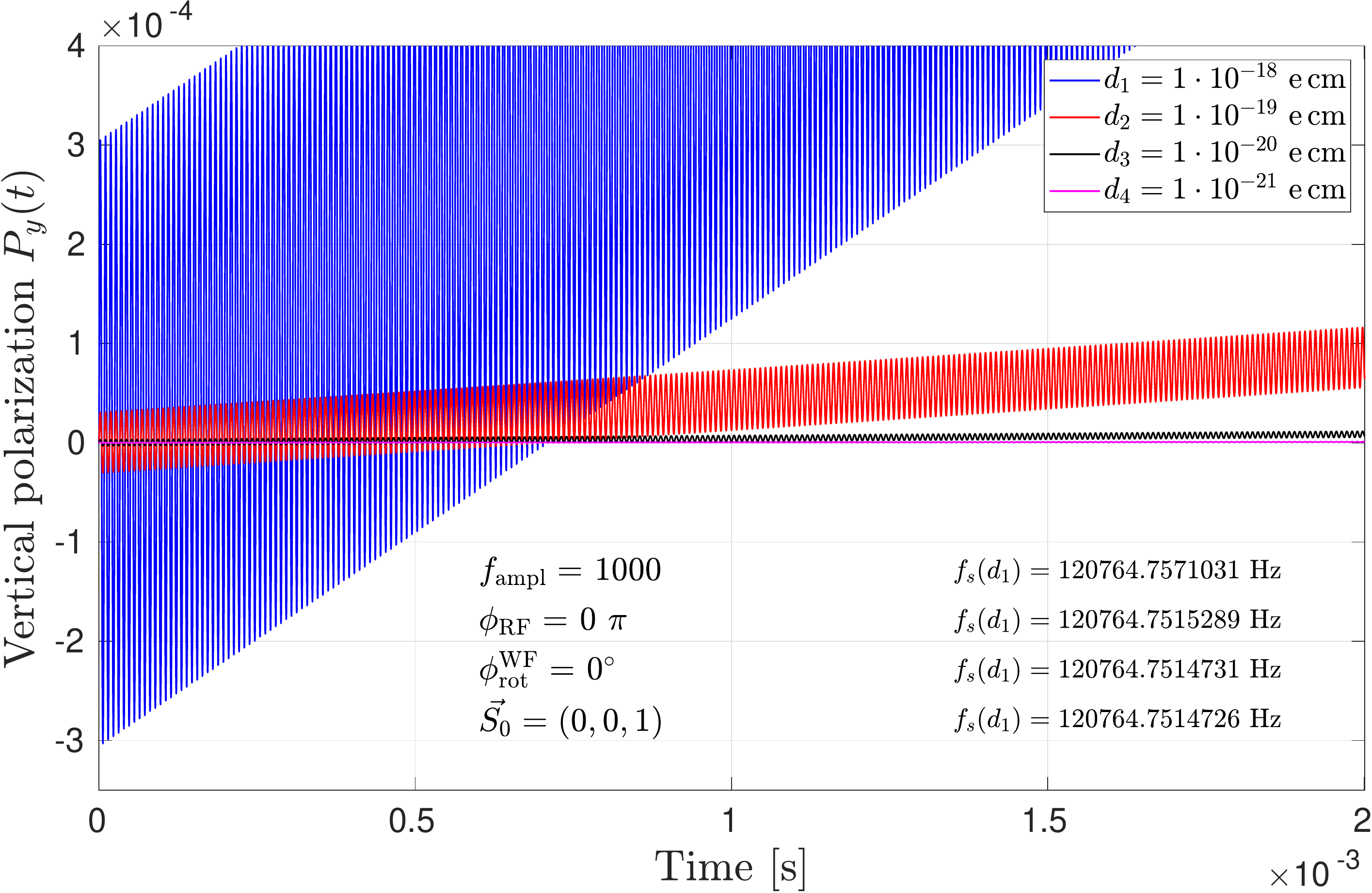}
\end{center}
\caption{\label{fig:model-calculation} Model calculation of the buildup of a vertical polarization component $P_y(t)$ for four different EDMs ranging from $\num{e-18}$ to $\num{e-21}$ and the conditions as indicated with an initial polarization $\vec P(t=0) = (0,0,1)$. In order to make things visible on this timescale,  a field amplification factor of $f_\text{ampl} = 1000$ has been used here to enhance the effects (see Eq.\,(\ref{eq:field-enhancement-factor})).}
\end{figure}

A model calculation of the polarization buildup, essentially showing only the very beginning of the polarization oscillation,  is shown in Fig.\,\ref{fig:model-calculation}. Actually, the term \textit{buildup} is meant here as an out-of-plane rotation of the initial purely in-plane polarization due to the presence of either an EDM and/or unwanted MDM rotations due to field imperfections and a non-ideal closed orbit in the ring, because ideally, the magnitude of the polarization $|\vec P(t)|$ remains constant. 

The model calculation reflects the situation of an ideal COSY ring with stored deuterons at $p_d = \SI{970}{MeV \per c}$ ($G_d = -0.143$,  $\gamma=1.126$). The spin precession frequency under these conditions amounts to 
\begin{equation}
 f_s =  f_{\text{rev}} (\gamma G \pm K)  \quad \approx \SI{120.765}{kHz} \text{ for } K=0\,.
\end{equation}
As can be seen in Fig.\,\ref{fig:model-calculation} the oscillation due to the spin precession is vastly faster than the change $\dd P_y(t)/\dd t$ due to the EDM. The electric field integral assumed in the model calculation is 
\begin{equation}
f_\text{ampl} \times \int E_\text{WF} \cdot  \dd \ell \approx \SI{2200}{kV} \quad \text{(w/o ferrites)}, \quad \text{ where} \quad f_\text{ampl} = 1000\,.
\label{eq:field-enhancement-factor}
\end{equation}
The assumed electric and magnetic fields are by a factor 1000 larger than the fields of the real WF operated (without ferrites) at an input  power of $\SI{1}{kW}$\,\cite{Slim:2016pim}. 

\subsection{Measurements of EDM-like polarization buildup}

Alternatively, $\varepsilon^\text{EDM}$ can be determined from the measured initial slopes $\dot p_y(t)|_{t=0}$ of the polarization buildup through a variation of the RF phase $\phi_\text{RF}$ using the phase-lock (see Sec.\,\ref{sec:phase-locking-the-spin-precession}):
\begin{equation}
  \varepsilon^\text{EDM} = \frac{\left. \dot p_y(t) \right|_{t=0}}{a\,\cos\phi_\text{RF}} \cdot \frac{1}{\Omega^\text{rev}} \,.
\label{eq:resonance-strength-from-phi-variation}
\end{equation}

\begin{figure}[bt]
 \def\tilt{38.23}
 \centering
\tdplotsetmaincoords{65}{250} 
\begin{tikzpicture} [scale=7, tdplot_main_coords, axis/.style={->,blue,thick}, 
vector/.style={-stealth,red,very thick}, 
vector guide/.style={dashed,red,thick}]
\tikzstyle{every node}=[font=\small]

%standard tikz coordinate definition using x, y, z coords
\coordinate (O) at (0,0,0);

%tikz-3dplot coordinate definition using x, y, z coords

\pgfmathsetmacro{\ax}{0.7}
\pgfmathsetmacro{\ay}{0.6}
\pgfmathsetmacro{\az}{0.3}

\pgfmathsetmacro{\aX}{0.84}
\pgfmathsetmacro{\aY}{0.72}
\pgfmathsetmacro{\aZ}{0.0}

\coordinate (P) at (\ax,\ay,\az);

%draw axes
\draw[axis] (0,0,0) -- (1,0,0)     node[anchor= west, yshift = 0.2cm]{$z$};
\draw[axis] (0,0,0) -- (0,0.8,0)   node[anchor=north west]{$x$};
\draw[axis] (0,0,0) -- (0,0,0.5)   node[anchor=west]{$y$};

%draw a vector from O to P
\draw[very thick, -stealth] (O) -- (P) node[anchor = east, yshift = 0.3cm, xshift = +1cm ]{$\vec{P}(t)$};

%draw guide lines to components
\draw[-stealth, very thick]   (O) -- (\ax,\ay,0) node[anchor = north, xshift = -0.4cm, yshift = 0cm ]{$\vec{P}_{xz}(t)$};
\draw[red, very thick, -stealth] (\ax,\ay,0) -- (P) node[anchor = east, yshift = -0.3cm,]{$\vec{P}_{y}(t)$};
\draw[gray, very thin] (\ax,\ay,0) -- (0,\ay,0);
\draw[gray, very thin] (\ax,\ay,0) -- (\ax,0,0);

\tdplotdefinepoints(0,0,0)(\ax, \ay, 0)(\ax, \ay, \az)
\tdplotdrawpolytopearc[very thick, red, -stealth]{0.5}{red, below, yshift=+0.5cm, xshift = -0.45cm}{\small $\alpha(t)$}

\tdplotdefinepoints(0,0,0)(0,0.5,0)(\ax,\ay,0)
% \tdplotdrawpolytopearc[thick, caribbeangreen, -stealth]{0.4}{caribbeangreen, east, yshift=+0.1 cm, xshift = -0.4cm}{$\phi_{S^x_0}$}
\tdplotdrawpolytopearc[thick, -stealth]{0.4}{anchor = east, yshift=+0.1 cm, xshift = -0.4cm}{\small $\phi_\text{RF}$}
\end{tikzpicture}
% \vspace{-0.2cm}
% \caption{\label{fig:inclination-angle-alpha} \small $\vec P_y(t)$ and inclination angle $\alpha$.% (see also Fig.\,\ref{fig:definition-phi_SX}).
% }
\caption{\label{fig:pydotstuff} The deuteron spins are precessing in the horizontal ($xz$) plane and the RF Wien filter is running on the corresponding frequency with a certain RF phase $\phi_\text{RF}$ that is maintained using the phase-locking system, discussed in Sec.\,\ref{sec:phase-locking-the-spin-precession}. }
\end{figure}
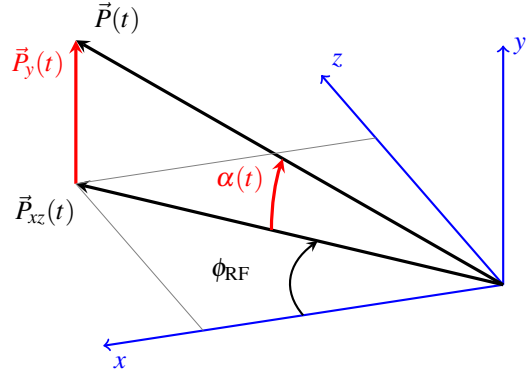

One can show that the evaluation of $\varepsilon^\text{EDM}$ from Eqs.\,(\ref{eq:resonance-strength}) and (\ref{eq:resonance-strength-from-phi-variation}) is equivalent if $|\vec P|= 1$, \textit{i.e.}, $\dot p_y(t) = \dot \alpha(t)$. This situation is indicated in Fig.\,\ref{fig:pydotstuff}.

\begin{figure}[hbt]
\begin{center}
\subfigure[\label{fig:build-up-3anglesa} \small $\dot \alpha$ for $\phi^\text{WF}_\text{rot} = -\SI{1}{\degree}$, \SI{0}{\degree}, $+\SI{1}{\degree}$ and $\chi^{\text{Snake}}_\text{rot} = 0$.]
{\includegraphics[height=0.23\textheight]{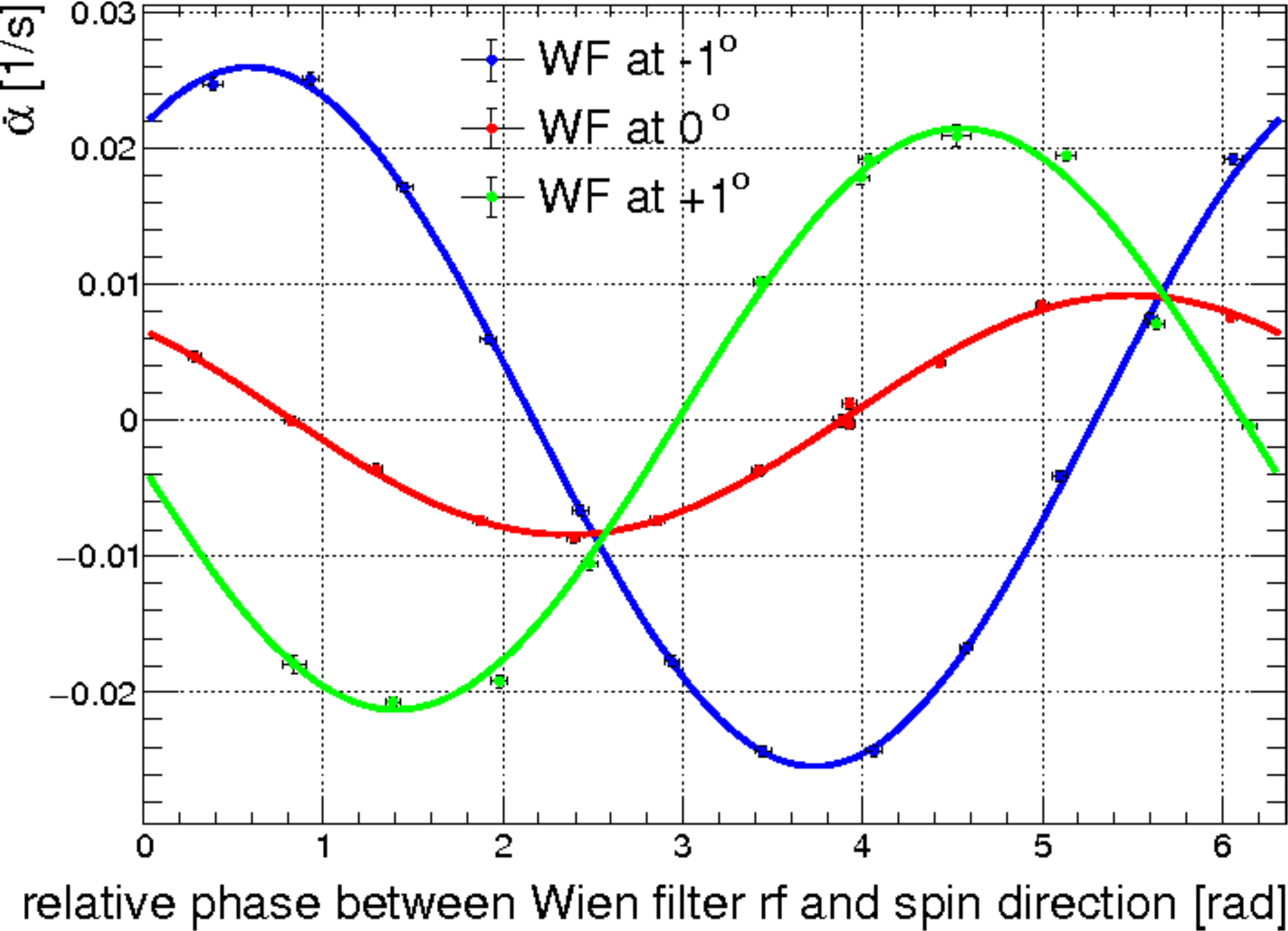}}
\hspace{0.2cm}
\subfigure[\label{fig:build-up-3anglesb} \small $\dot \alpha$ for $\chi^{\text{Snake}}_\text{rot} = -1$, 0, $+\SI{1}{\degree}$ and $\phi^\text{WF}_\text{rot} =0$.]
{\includegraphics[height=0.23\textheight]{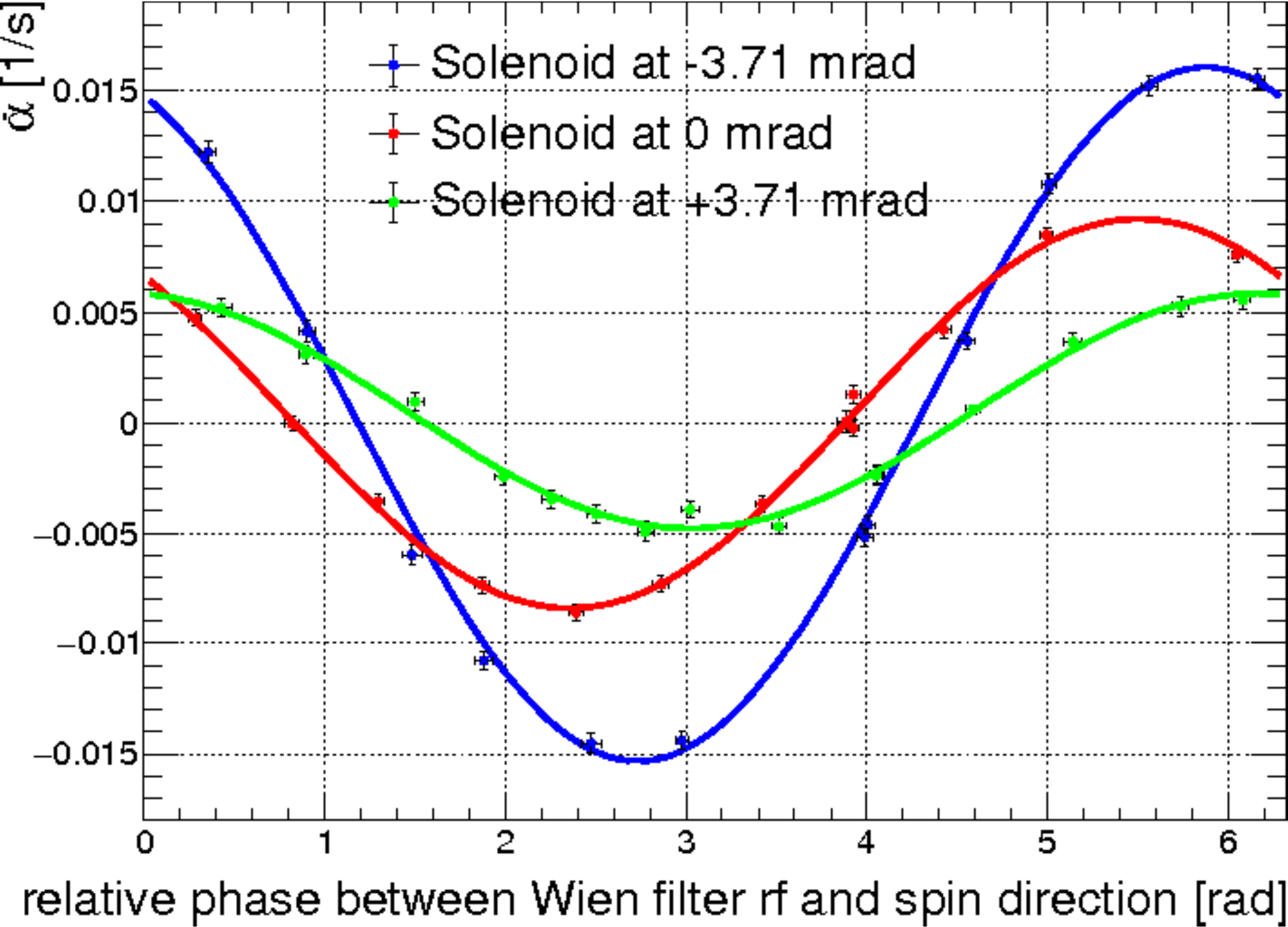}}\\
\end{center}
\caption{\label{fig:build-up-3angles} \small Rate of the out-of-plane rotation angle $\dot \alpha(t)|_{t=0}$ as function of the Wien filter RF phase $\phi_\text{RF}$ for two situations. In panel (a), only the RF Wien filter is rotated around the beam axis, and in (b) only the Siberian snake solenoid in the opposite straight section of COSY (see Fig.\,\ref{fig:COSY-landscape}) rotates the spins around the beam axis.}
\end{figure}

The first measurements of EDM-like buildup signals by JEDI are shown in Fig.\,\ref{fig:build-up-3angles}. Both plots show the rate of the out-of-plane rotation angle $\dot \alpha(t)|_{t=0}$ as function of the Wien filter RF phase ($\phi_\text{RF}$) for different rotations of the RF Wien filter around the beam axis ($\phi^\text{WF}_\text{rot}$) and different spin rotations in the Snake solenoid ($\chi^{\text{Snake}}_\text{rot}$).
For these measurements, the $B$ field of RF Wien filter is oriented normal to the ring plane (along $\vec e_y$). The RF Wien filter was operated at $f_\text{WF} = \SI{871}{kHz}$.  Variations of $\phi^\text{WF}_\text{rot}$ and $\chi^{\text{Snake}}_\text{rot}$ affect the pattern of observed initial slopes $\dot \alpha$. 
During the measurements shown in Fig.\,\ref{fig:build-up-3angles}, the magnets of the electron cooler were switched off altogether on flattop to reduce unwanted spin precessions of the stored particles in the cooler magnets. 

Further details about this type of EDM measurement are discussed in the contribution to these proceedings by Alexander Nass\,\cite{spin2018:Alexander.Nass}. As next step the first EDM production run using COSY is scheduled for Nov.-Dec.\ 2018.

%---------------------------------------------------------------------------------------------------------------------------------
\section{Axion-EDM search using storage ring}
%---------------------------------------------------------------------------------------------------------------------------------
The motivation to search for oscillating axion-EDMs using storage rings is derived from a recent paper by Graham and Rajendran\,\cite{PhysRevD.84.055013}. An oscillating axion field couples to gluons and induces an oscillating EDM in hadronic particles. The measurement principle relies on the fact that when the oscillating EDM resonates with the particle $g-2$ precession frequency in the storage ring, the EDM precession can be accumulated. Furthermore, due to the strong effective electric field (from $\vec v \times \vec B$), the sensitivity is improved significantly. Limits for axion-gluon couplings to oscillating EDMs are discussed in Ref.\,\cite{Chang:2017ruk}.
 
Without any additional equipment, a measurement of axion-like oscillating EDMs can be realized in the magnetic storage ring COSY. A proposal for a test beam time\protect\footnote{Available from \url{http://collaborations.fz-juelich.de/ikp/jedi/public_files/proposals/Axion_Search_at_COSY.pdf}} has been accepted by the COSY Beam time Advisory Committee. A first experiment is scheduled for the first half of 2019.

%----------------------------------------------------------------------------------------------------------
\section{Summary}
%----------------------------------------------------------------------------------------------------------
The search for charged particle EDMs offers a new window to disentangle sources of $CP$ violation, and to possibly explain the matter-antimatter asymmetry of the Universe. The JEDI collaboration is making steady progress in the field of spin dynamics of relevance to future searches of EDMs. For these investigations, COSY remains to be a unique facility. 

The first direct JEDI deuteron EDM measurement at COSY is well underway. A first run took place in Nov.-Dec.\ 2018, a second run shall take place at the end of 2019. The anticipated deuteron EDM sensitivity of the measurements is about $\num{e-18}$ to $\SI{e-20}{e.cm}$.

There is a strong interest of the high-energy physics community in storage ring searches for the EDM of protons and light nuclei as part of physics program of the post-LHC era. In the framework of the recently formed CPEDM Collaboration that evolved out of JEDI, the design of a \SI{30}{MeV} all-electric  EDM prototype storage ring is being prepared. Possible hosts are CERN or COSY. 

\acknowledgments
This work is supported by an ERC Advanced-Grant of the European Union (proposal number 694340).

%----------------------------------------------------------------------------------------------------------------------------------
\bibliographystyle{JHEP}
\bibliography{/home/frank/Dokumente/Documents/LITERATUR/spintunemapping_23.01.2017}

\providecommand{\href}[2]{#2}\begingroup\raggedright\begin{thebibliography}{10}

\bibitem{Sakharov:1967dj}
A.~D. Sakharov, \emph{{Violation of CP Invariance, c Asymmetry, and Baryon
  Asymmetry of the Universe}},
  \href{https://doi.org/10.1070/PU1991v034n05ABEH002497}{\emph{Pisma Zh. Eksp.
  Teor. Fiz.} {\bfseries 5} (1967) 32}.

\bibitem{PhysRevLett.110.101802}
{\scshape LHCb} collaboration, \emph{{Observation of
  ${D}^{0}\mathbf{\ensuremath{-}}{\overline{D}}^{0}$ Oscillations}},
  \href{https://doi.org/10.1103/PhysRevLett.110.101802}{\emph{Phys. Rev. Lett.}
  {\bfseries 110} (2013) 101802}.

\bibitem{Polycarpo:2012yh}
E.~Polycarpo, \emph{{D0 Mixing and CP Violation in D Decays}},  in
  \emph{{Proceedings, 32nd International Symposium on Physics in Collision (PIC
  2012): Strbske Pleso, Slovakia, September 12-15, 2012}}, pp.~219--228, 2012,
  \href{https://arxiv.org/abs/1212.6341}{{\ttfamily 1212.6341}}.

\bibitem{Bennett:2003bz}
{\scshape WMAP} collaboration, \emph{{First year Wilkinson Microwave Anisotropy
  Probe (WMAP) observations: Preliminary maps and basic results}},
  \href{https://doi.org/10.1086/377253}{\emph{Astrophys. J. Suppl.} {\bfseries
  148} (2003) 1} [\href{https://arxiv.org/abs/astro-ph/0302207}{{\ttfamily
  astro-ph/0302207}}].

\bibitem{Barger:2003zg}
V.~Barger, J.~P. Kneller, H.-S. Lee, D.~Marfatia and G.~Steigman,
  \emph{{Effective number of neutrinos and baryon asymmetry from BBN and
  WMAP}}, \href{https://doi.org/10.1016/S0370-2693(03)00800-1}{\emph{Phys.
  Lett.} {\bfseries B566} (2003) 8}
  [\href{https://arxiv.org/abs/hep-ph/0305075}{{\ttfamily hep-ph/0305075}}].

\bibitem{Bernreuther:2002uj}
W.~Bernreuther, \emph{{CP violation and baryogenesis}}, {\emph{Lect. Notes
  Phys.} {\bfseries 591} (2002) 237}
  [\href{https://arxiv.org/abs/hep-ph/0205279}{{\ttfamily hep-ph/0205279}}].

\bibitem{Khriplovich:1997ga}
I.~B. Khriplovich and S.~K. Lamoreaux, \emph{{CP violation without strangeness:
  Electric dipole moments of particles, atoms, and molecules}}. Berlin,
  Germany: Springer (1997) 230 p, 1997.

\bibitem{Afach:2015sja}
J.~M. Pendlebury, S.~Afach, N.~J. Ayres, C.~A. Baker, G.~Ban, G.~Bison et~al.,
  \emph{Revised experimental upper limit on the electric dipole moment of the
  neutron}, \href{https://doi.org/10.1103/PhysRevD.92.092003}{\emph{Phys. Rev.
  D} {\bfseries 92} (2015) 092003}.

\bibitem{roberts-marciano:2010}
B.~L. Roberts and W.~J. Marciano, eds., \emph{Lepton Dipole Moments}, vol.~20
  of \emph{Advanced Series on Directions in High Energy Physics}. World
  Scientific, 2010.

\bibitem{Bsaisou:2015:1}
J.~Bsaisou, J.~de~Vries, C.~Hanhart, S.~Liebig, U.-G. Mei{\ss}ner, D.~Minossi
  et~al., \emph{Nuclear electric dipole moments in chiral effective field
  theory}, \href{https://doi.org/10.1007/JHEP03(2015)104}{\emph{Journal of High
  Energy Physics} {\bfseries 2015} (2015) 1}.

\bibitem{Bsaisou:2015:2}
J.~Bsaisou, J.~de~Vries, C.~Hanhart, S.~Liebig, U.-G. Mei{\ss}ner, D.~Minossi
  et~al., \emph{Erratum to: Nuclear electric dipole moments in chiral effective
  field theory}, \href{https://doi.org/10.1007/JHEP05(2015)083}{\emph{Journal
  of High Energy Physics} {\bfseries 2015} (2015) 1}.

\bibitem{Guo:2012vf}
F.-K. Guo and U.-G. Meissner, \emph{{Baryon electric dipole moments from strong
  CP violation}}, \href{https://doi.org/10.1007/JHEP12(2012)097}{\emph{JHEP}
  {\bfseries 12} (2012) 097} [\href{https://arxiv.org/abs/1210.5887}{{\ttfamily
  1210.5887}}].

\bibitem{Flambaum:1985gv}
V.~V. Flambaum, I.~B. Khriplovich and O.~P. Sushkov, \emph{{On the $P$ and $T$
  Nonconserving Nuclear Moments}},
  \href{https://doi.org/10.1016/0375-9474(86)90331-3}{\emph{Nucl. Phys.}
  {\bfseries A449} (1986) 750}.

\bibitem{Flambaum:1985ty}
V.~V. Flambaum, I.~B. Khriplovich and O.~P. Sushkov, \emph{{Limit on the
  constant of T non-conserving Nucleon-Nucleon interaction}},
  \href{https://doi.org/10.1016/0370-2693(85)90908-6}{\emph{Phys. Lett.}
  {\bfseries 162B} (1985) 213}.

\bibitem{ACME2014}
J.~Baron, W.~C. Campbell, D.~DeMille, J.~M. Doyle, G.~Gabrielse, Y.~V. Gurevich
  et~al., \emph{Order of magnitude smaller limit on the electric dipole moment
  of the electron},
  \href{https://doi.org/10.1126/science.1248213}{\emph{Science} {\bfseries 343}
  (2014) 269}
  [\href{https://arxiv.org/abs/http://science.sciencemag.org/content/343/6168/269.full.pdf}{{\ttfamily
  http://science.sciencemag.org/content/343/6168/269.full.pdf}}].

\bibitem{PhysRevD.80.052008}
{\scshape Muon (g-2)} collaboration, \emph{Improved limit on the muon electric
  dipole moment}, \href{https://doi.org/10.1103/PhysRevD.80.052008}{\emph{Phys.
  Rev. D} {\bfseries 80} (2009) 052008}.

\bibitem{INAMI200316}
K.~Inami, K.~Abe, K.~Abe, R.~Abe, T.~Abe, I.~Adachi et~al., \emph{Search for
  the electric dipole moment of the $\tau$ lepton},
  \href{https://doi.org/https://doi.org/10.1016/S0370-2693(02)02984-2}{\emph{Physics
  Letters B} {\bfseries 551} (2003) 16 }.

\bibitem{PhysRevD.23.814}
L.~Pondrom, R.~Handler, M.~Sheaff, P.~T. Cox, J.~Dworkin, O.~E. Overseth
  et~al., \emph{New limit on the electric dipole moment of the
  $\ensuremath{\Lambda}$ hyperon},
  \href{https://doi.org/10.1103/PhysRevD.23.814}{\emph{Phys. Rev. D} {\bfseries
  23} (1981) 814}.

\bibitem{PhysRevLett.91.212303}
V.~F. Dmitriev and R.~A. Sen'kov, \emph{Schiff moment of the mercury nucleus
  and the proton dipole moment},
  \href{https://doi.org/10.1103/PhysRevLett.91.212303}{\emph{Phys. Rev. Lett.}
  {\bfseries 91} (2003) 212303}.

\bibitem{PhysRevLett.116.161601}
B.~Graner, Y.~Chen, E.~G. Lindahl and B.~R. Heckel, \emph{Reduced limit on the
  permanent electric dipole moment of $^{199}\mathrm{Hg}$},
  \href{https://doi.org/10.1103/PhysRevLett.116.161601}{\emph{Phys. Rev. Lett.}
  {\bfseries 116} (2016) 161601}.

\bibitem{PhysRevLett.86.22}
M.~A. Rosenberry and T.~E. Chupp, \emph{Atomic electric dipole moment
  measurement using spin exchange pumped masers of $^{129}{Xe}$ and
  $^{3}{He}$}, \href{https://doi.org/10.1103/PhysRevLett.86.22}{\emph{Phys.
  Rev. Lett.} {\bfseries 86} (2001) 22}.

\bibitem{PhysRevLett.115.094801}
{\scshape JEDI} collaboration, \emph{New method for a continuous determination
  of the spin tune in storage rings and implications for precision
  experiments},
  \href{https://doi.org/10.1103/PhysRevLett.115.094801}{\emph{Phys. Rev. Lett.}
  {\bfseries 115} (2015) 094801}.

\bibitem{PhysRevSTAB.17.052803}
Z.~Bagdasarian, S.~Bertelli, D.~Chiladze, G.~Ciullo, J.~Dietrich, S.~Dymov
  et~al., \emph{Measuring the polarization of a rapidly precessing deuteron
  beam}, \href{https://doi.org/10.1103/PhysRevSTAB.17.052803}{\emph{Phys. Rev.
  ST Accel. Beams} {\bfseries 17} (2014) 052803}.

\bibitem{PhysRevLett.117.054801}
{\scshape JEDI} collaboration, \emph{How to reach a thousand-second in-plane
  polarization lifetime with $0.97\text{-}\mathrm{GeV}/c$ deuterons in a
  storage ring},
  \href{https://doi.org/10.1103/PhysRevLett.117.054801}{\emph{Phys. Rev. Lett.}
  {\bfseries 117} (2016) 054801}.

\bibitem{PhysRevLett.119.014801}
{\scshape JEDI} collaboration, \emph{Phase locking the spin precession in a
  storage ring},
  \href{https://doi.org/10.1103/PhysRevLett.119.014801}{\emph{Phys. Rev. Lett.}
  {\bfseries 119} (2017) 014801}.

\bibitem{PhysRevAccelBeams.21.042002}
{\scshape JEDI} collaboration, \emph{Phase measurement for driven spin
  oscillations in a storage ring},
  \href{https://doi.org/10.1103/PhysRevAccelBeams.21.042002}{\emph{Phys. Rev.
  Accel. Beams} {\bfseries 21} (2018) 042002}.

\bibitem{PhysRevAccelBeams.20.072801}
{\scshape JEDI} collaboration, \emph{Spin tune mapping as a novel tool to probe
  the spin dynamics in storage rings},
  \href{https://doi.org/10.1103/PhysRevAccelBeams.20.072801}{\emph{Phys. Rev.
  Accel. Beams} {\bfseries 20} (2017) 072801}.

\bibitem{Brantjes201249}
N.~Brantjes, V.~Dzordzhadze, R.~Gebel, F.~Gonnella, F.~Gray, D.~van~der Hoek
  et~al., \emph{Correcting systematic errors in high-sensitivity deuteron
  polarization measurements},
  \href{https://doi.org/https://doi.org/10.1016/j.nima.2011.09.055}{\emph{Nuclear
  Instruments and Methods in Physics Research Section A: Accelerators,
  Spectrometers, Detectors and Associated Equipment} {\bfseries 664} (2012) 49
  }.

\bibitem{Grigoryev:2018ipo}
K.~Grigoryev, F.~Rathmann, A.~Stahl and H.~Str\"oher, \emph{{Electrostatic
  deflector studies using small prototypes}},
  \href{https://arxiv.org/abs/1812.07954}{{\ttfamily 1812.07954}}.

\bibitem{Trinkel:2017}
F.~Trinkel, \emph{{Development of a Rogowski coil Beam Position Monitor for
  Electric Dipole Moment measurements at storage rings}}, Ph.D. thesis, RWTH
  Aachen University, 2017.

\bibitem{spin2018:Alexander.Nass}
A.~Nass, \emph{{Commissioning of the RF Wien filter for a first deuteron EDM
  measurement at COSY/J\"ulich}},  in \emph{{Proceedings, 23rd International
  Spin Physics Symposium: Ferrara, Italy, 10-14 September, 2018}}, 2018.

\bibitem{spin2018:Fabian.Mueller}
F.~M\"uller, \emph{{Measurement of dC Vector Analyzing Power and Cross Sections
  at COSY for EDM Polarimetry}},  in \emph{{Proceedings, 23rd International
  Spin Physics Symposium: Ferrara, Italy, 10-14 September, 2018}}, 2018.

\bibitem{Pretz2019}
J.~Pretz and F.~M{\"u}ller, \emph{Extraction of azimuthal asymmetries using
  optimal observables},
  \href{https://doi.org/10.1140/epjc/s10052-019-6580-3}{\emph{The European
  Physical Journal C} {\bfseries 79} (2019) 47}.

\bibitem{Albers:2004iw}
D.~Albers et~al., \emph{{A Precision measurement of pp elastic scattering
  cross-sections at intermediate energies}},
  \href{https://doi.org/10.1140/epja/i2004-10011-3}{\emph{Eur. Phys. J.}
  {\bfseries A22} (2004) 125}.

\bibitem{Adam:2004ch}
{\scshape WASA-at-COSY} collaboration, \emph{{Proposal for the wide angle
  shower apparatus (WASA) at COSY-Julich: WASA at COSY}},
  \href{https://arxiv.org/abs/nucl-ex/0411038}{{\ttfamily nucl-ex/0411038}}.

\bibitem{Bargholtz:2008aa}
{\scshape CELSIUS/WASA} collaboration, \emph{{The WASA Detector Facility at
  CELSIUS}}, \href{https://doi.org/10.1016/j.nima.2008.06.011}{\emph{Nucl.
  Instrum. Meth.} {\bfseries A594} (2008) 339}
  [\href{https://arxiv.org/abs/0803.2657}{{\ttfamily 0803.2657}}].

\bibitem{spin2018:Dito.Shergelashvili}
D.~Shergelashvili, D.~Mchedlishvili, F.~M\"uller and I.~Keshelashvili,
  \emph{{Development of LYSO detector modules for a charge-particle EDM
  polarimeter}},  in \emph{{Proceedings, 23rd International Spin Physics
  Symposium: Ferrara, Italy, 10-14 September, 2018}}, 2018.

\bibitem{spin2018:Kolya.Nikolaev}
N.~Nikolaev, F.~Rathmann, A.~Saleev and A.~Silenko, \emph{{Gravity and Spin
  Dynamics for the EDM Search Experiments}},  in \emph{{Proceedings, 23rd
  International Spin Physics Symposium: Ferrara, Italy, 10-14 September,
  2018}}, 2018.

\bibitem{doi:10.1063/1.4967465}
V.~Anastassopoulos, S.~Andrianov, R.~Baartman, S.~Baessler, M.~Bai, J.~Benante
  et~al., \emph{A storage ring experiment to detect a proton electric dipole
  moment}, {\emph{Review of Scientific Instruments} {\bfseries 87} (2016)
  115116}.

\bibitem{spin2018:Sig.Martin}
A.~Lehrach, S.~Martin and R.~M. Talman, \emph{{Design of a Prototype EDM
  Storage Ring}},  in \emph{{Proceedings, 23rd International Spin Physics
  Symposium: Ferrara, Italy, 10-14 September, 2018}}, 2018.

\bibitem{Rathmann:2013rqa}
F.~Rathmann, A.~Saleev and N.~N. Nikolaev, \emph{{The search for electric
  dipole moments of light ions in storage rings}},
  \href{https://doi.org/10.1088/1742-6596/447/1/012011}{\emph{J. Phys. Conf.
  Ser.} {\bfseries 447} (2013) 012011}.

\bibitem{PhysRevSTAB.16.114001}
W.~M. Morse, Y.~F. Orlov and Y.~K. Semertzidis, \emph{rf wien filter in an
  electric dipole moment storage ring: The ``partially frozen spin'' effect},
  \href{https://doi.org/10.1103/PhysRevSTAB.16.114001}{\emph{Phys. Rev. ST
  Accel. Beams} {\bfseries 16} (2013) 114001}.

\bibitem{Slim:2016pim}
J.~Slim et~al., \emph{{Electromagnetic Simulation and Design of a Novel
  Waveguide RF Wien Filter for Electric Dipole Moment Measurements of Protons
  and Deuterons}},
  \href{https://doi.org/10.1016/j.nima.2016.05.012}{\emph{Nucl. Instrum. Meth.}
  {\bfseries A828} (2016) 116}
  [\href{https://arxiv.org/abs/1603.01567}{{\ttfamily 1603.01567}}].

\bibitem{Slim:2017bic}
J.~Slim, F.~Rathmann and D.~Heberling, \emph{{Computational framework for
  particle and spin simulations based on the stochastic Galerkin method}},
  \href{https://doi.org/10.1103/PhysRevE.96.063301}{\emph{Phys. Rev.}
  {\bfseries E96} (2017) 063301}
  [\href{https://arxiv.org/abs/1707.09274}{{\ttfamily 1707.09274}}].

\bibitem{Augustin2012}
F.~Augustin and P.~Rentrop, \emph{Stochastic galerkin techniques for random
  ordinary differential equations},
  \href{https://doi.org/10.1007/s00211-012-0466-8}{\emph{Numerische Mathematik}
  {\bfseries 122} (2012) 399}.

\bibitem{Slim:2016dct}
J.~Slim, F.~Rathmann, A.~Nass, H.~Soltner, R.~Gebel, J.~Pretz et~al.,
  \emph{{Polynomial Chaos Expansion method as a tool to evaluate and quantify
  field homogeneities of a novel waveguide RF Wien Filter}},
  \href{https://doi.org/10.1016/j.nima.2017.03.040}{\emph{Nucl. Instrum. Meth.}
  {\bfseries A859} (2017) 52}
  [\href{https://arxiv.org/abs/1612.09235}{{\ttfamily 1612.09235}}].

\bibitem{PhysRevD.84.055013}
P.~W. Graham and S.~Rajendran, \emph{Axion dark matter detection with cold
  molecules}, \href{https://doi.org/10.1103/PhysRevD.84.055013}{\emph{Phys.
  Rev. D} {\bfseries 84} (2011) 055013}.

\bibitem{Chang:2017ruk}
S.~P. Chang, S.~Haciomeroglu, O.~Kim, S.~Lee, S.~Park and Y.~K. Semertzidis,
  \emph{{Axion dark matter search using the storage ring EDM method}},
  {\emph{PoS} {\bfseries PSTP2017} (2018) 036}
  [\href{https://arxiv.org/abs/1710.05271}{{\ttfamily 1710.05271}}].

\end{thebibliography}\endgroup
%----------------------------------------------------------------------------------------------------------------------------------

%----------------------------------------------------------------------------------------------------------------------------------
\end{document}